\documentclass[fleqn,twoside]{article}
\usepackage[pdftex,colorlinks=true,urlcolor=black,filecolor=black,linkcolor=black,
            pdftitle={The Kasner Brane.},
            pdfauthor={Mark D. Roberts},
            pdfsubject={relativity,cosmology},
            pdfkeywords={kasner,exact solutions},
            pagebackref,pdfpagemode=None,bookmarksopen=true]{hyperref}
\usepackage{amssymb}
\newcommand{\bc}{\begin{center}}
\newcommand{\ec}{\end{center}}
\newcommand{\bi}{\begin{itemize}}     
\newcommand{\ei}{\end{itemize}}
\newcommand{\bd}{\begin{description}} 
\newcommand{\ed}{\end{description}}
\newcommand{\bn}{\begin{enumerate}}   
\newcommand{\en}{\end{enumerate}}
\newcommand{\be}{\begin{equation}}
\newcommand{\ee}{\end{equation}}
\newcommand{\ber}{\begin{eqnarray}}
\newcommand{\ear}{\end{eqnarray}}
\newcommand{\ba}{\begin{array}}
\newcommand{\ea}{\end{array}}
\newcommand{\bv}{\begin{verbatim}}
\newcommand{\ev}{\end{verbatim}}

\newcommand{\bx}{\square}

\newcommand{\de}{\delta}

\newcommand{\fr}{\frac}

\newcommand{\lb}{\label}

\newcommand{\n}{\nonumber\\}

\newcommand{\Order}{{\cal O}}

\newcommand{\pr}{{\cal P}}

\newcommand{\sq}{\sqrt}
\newcommand{\st}{\stackrel}

\begin{document}
\title{The Kasner Brane.}
\author{Mark D. Roberts,\\
\href{http://www.aei.mpg.de/}{Max Planck Institute for Gravitational Physics (Albert Einstein Institute)}\\
1 Am M\"uhlenberg,  Golm,  Germany D-14476\\
mdr@ihes.fr
}
\date{$12^{th}$ of November 2009}
\maketitle
\begin{abstract}
Solutions are found to field equations constructed from the
Pauli, Bach and Gauss-Bonnet quadratic tensors
to the Kasner and Kasner brane spacetimes in up to five dimensions.
A double Kasner space is shown to have a vacuum solution.
Brane solutions in which the bulk components of the Einstein tensor vanish are also looked at
and for four branes a solution similar to radiation Robertson-Walker spacetime is found.
Matter trapping of a test scalar field and a test perfect fluid
are investigated using energy conditions.
\end{abstract}
\href{http://arXiv.org/abs/gr-qc/0510005}
                      {\tt gr-qc/0510005}
{\scriptsize\tableofcontents}
\section{Introduction}
Randall and Sundrum \cite{RS} considered models where four dimensional spacetime
is multiplied by a conformal factor which has dependence on the fifth dimension.
Originally the four dimensional spacetime was taken to be flat and the dependence
on the fifth dimension such that overall the space was five dimensional anti-deSitter space.
Apart from flat four dimensional spacetime,
four dimensional G\"odel spacetime Barrow and Tsagas \cite{BT}
and four dimensional Bianchi type IX cosmologies
van den Hoogen, Coley and He \cite{HCH}
have been considered.
Here,  instead of four dimensional flat spacetime,  Kasner spacetime in various dimensions
is considered,  with a view to seeing what sort of field equations can be obeyed.
The field equations that are looked at are those involving quadratic curvature,
as such field equations often occur in fundamental theories such as string theories.
There is a choice of using field equations from Lagrangians involving the Ricci scalar squared,
the Ricci tensor squared, and the Riemann tensor squared,  or equivalently using the Pauli,
Bach and Gauss-Bonnet tensors.
The later has the advantage that interpretation of the tensors is more immediate.
The Pauli tensor \cite{pauli} comes from varying a Ricci scalar squared Lagrangian and
is defined by
\be
P_{ab}=2R_{;a;b}-2RR_{ab}+g_{ab}(R^2/2-2\bx R),
\label{pauli}
\ee
as it comes from varying the simplest quadratic scalar it is the most frequantly studied.
The Bach tensor \cite{bach} comes from varying the Weyl squared Lagrangian and
is defined by
\be
B_{a b}=2C_{a..b}^{~cd}R_{cd}+4C_{a..b;c;d}^{~cd},
\ee
it has several unusual properties such as:
trace-free,
conformal invariance,
asymptotically increasing (i.e. $\exp(+kr)$) spherically symmetric linearization,
faster-than-light linearization.
The Bach tensor often occurs in quantum theories where it cancels out divergent one-loop terms.
The Gauss-Bonnet invariant \cite{zwiebach} is
\be
GB=R_{cdef}R^{cdef}-4R_{cd}R^{cd}+R^2,
\label{GaussBonnet}
\ee
variation gives the Gauss-Bonnet tensor
\be
GB_{ab}=4R_{acde}R_b^{~cde}-8R_{cd}R_{~a~b}^{c~d}-8R_{ac}R^c_{~b}+4RR_{ab}-g_{ab}GB,
\label{Gb}
\ee
it vanishes in four dimensions,  but could be the dominant low energy quadratic curvature
contribution from string theory.
A scalar field has
\be
R_{ab}=2\phi_a\phi_b-g_{ab}V\left(\phi^2\right),
\label{sf}
\ee
and a perfect fluid
\be
G_{ab}=\left(\pr+\mu\right)U_aU_b+\pr g_{ab},~
R_{ab}=\left(\pr+\mu\right)U_aU_b+\frac{\pr-\mu}{2-d}g_{ab},
d\ne2.
\label{pf}
\ee
often the equation of state
\be
\pr=\left(\gamma-1\right)\mu,
\label{geqs}
\ee
is used.  Calculations were done using grtensor2/maple9 \cite{MPL}.
\section{Kasner spacetime in two dimensions.}
In two dimensions the line element is
\be
ds^2_2=-dt^2+t^{2p}dx^2,
\ee
$x$ is a Killing coordinate,  but not $t$.
The Weyl,  Bach and Gauss-Bonnet tensors either vanish or are not defined.
The Ricci tensor is given by
\be
R_{a b}=
\fr{p(p-1)}{t^2}
\left(
\begin{array}{c c}
t^{2p} & 0 \n 0& -1
\end{array}
\right).
\ee
The Pauli tensor is given by
\be
P_{a b}=
\fr{2p(p-1)(p+3)}{t^4}
\left(
\begin{array}{c c}
-(p-4)t^{2p} & 0 \n 0& p
\end{array}
\right),
\label{pd2}
\ee
so that when $p=-3$ there is a non-flat solution to the vacuum Pauli equations,
this contrasts to the $p=1$ case where the Ricci tensor vanishes but the spacetime is flat.
The curvature invariants are
\be
R=\fr{2p(p-1)}{t^2},~~~
RicciSq=\fr{1}{2}R^2,~~~
RiemSq=R^2,
\ee
when $p=1$ the spacetime is flat,  otherwise there is one degree of freedom.
For a vector field
\ber
&v^a\equiv\de^a_t,~~~
norm[v]=-1,~~~
v_a=[0,-1],~~~
\dot{v}^a=0,~~~
\st{v}{\Theta}=\fr{p}{t},\n
&\omega_{ab}=0,~~~
\sigma=\fr{2p}{3t},~~~
\sigma_{ab}=\de^{xx}_{ab}\fr{2}{3}t^{(2p-1)}.
\ear
\section{Kasner spacetime in three dimensions.}
The line element is
\be
ds^2=-dt^2+t^{2p_1}dx^2+t^{2p_2}dy^2,
\ee
The Weyl,  Bach and Gauss-Bonnet tensors either vanish or are not defined.
The Ricci tensor,  scalar and Pauli tensor up to symmetry in $x$ and $y$ are:
\ber
&R_{xx}=(p_1+p_2-1)t^{(2p_1-2)},~~~
R_{tt}=-\fr{1}{t^2}(-p_1-p_2+p_1^2+p_2^2),\n
&R=\fr{2}{t^2}(p_1^2+p_2^2+2p_1p_2-p_1-p_2),\\
&P_{xx}=\fr{R}{t^2}(p_1-p_1^2-p_1p_2+p_2^2-5p_2+12),~~~
P_{tt}=\fr{R}{t^2}(3p_1+p_1^2+3p_2+p_2^2-p_1p_2).\nonumber
\ear
No solutions for this line element have been found at all.
The traditional constraints $p_1+p_2=1,~p_1^2+p_2^2=1$,
here reduces to $p_1=0,~p_2=1$,  which gives vanishing $RiemSq$.
\section{Kasner spacetime in four dimensions.}
The metric is
\be
ds^2=-dt^2+t^{2p_1}dx^2+t^{2p_2}dy^2+t^{2p_3}dz^2,
\lb{kasnerle}
\ee
Let
\be
b\equiv p_1+p_2+p_3,~~~
a^2\equiv p_1^2+p_2^2+p_3^2,~~~
c\equiv p_1p_2+p_1p_3+p_2p_3,
\lb{ab}
\ee
then
\be
2c=b^2-a^2.
\lb{b2a2}
\ee
The Ricci tensor is given by
\be
R_{xx}=(b-1)p_1t^{(2p_1-2)},~~~
R_{tt}=(b-a^2)t^{-2},
\lb{kricci}
\ee
where the $y$ and $z$ components are given by symmetry.
The Bach tensor is
\ber
B_{xx}&=&\fr{t^{p_1}}{3t^4}(2a^2-3+2b-b^2)(3a^2-b^2-4b-4p_1^2+12p_1+4p_2p_3),\n
B_{tt}&=&\fr{1}{3t^4}(2a^2-3+2b-b^2)(3a^2-b^2).
\lb{kbach}
\ear
For the vector field
\be
v^a=\delta^a_t,
\lb{vec}
\ee
the acceleration,  vorticity,  and magnetic part of the Weyl tensor all vanish,  but
the expansion is
\be
\st{v}{\Theta}=\fr{b}{t},
\lb{expsc}
\ee
the shear tensor is
\be
\sigma_{xx}=-\fr{t^{2p_1-1}}{3}(-2p_1+p_2+p_3),~~~
\sigma_{tt}=0,
\lb{sheart}
\ee
the shear scalar is
\be
\sigma=\fr{\sq{3a^2-b^2}}{\sq{3}t}.
\lb{shears}
\ee
When $a^2=b=1$ both the Ricci tensor and the Bach tensor vanish,
and the curvature invariants are
\be
Ricciscalar=0,~~~
RicciSq=0,~~~
WeylSq=RiemSq=-\fr{16p_3^2(p_3-1)}{t^4},
\ee
this is the traditional solution;
however the Bach tensor also vanishes when
\be
2a^2-3+2b-b^2=0~~~
{\rm or}~~~
b=1\pm\sq{2a^2-2},
\lb{bsq}
\ee
and this combination of constants has other solutions.
For example:
the {\it expansion-free case}
when $b=0~\&~2a^2=3$, \ref{expsc} shows the expansion vanishes
and \ref{shears} shows non-vanishing shear scalar,
this has curvature products
\ber
&R=\fr{3}{2t^2},~~~
RicciSq=\fr{15}{4t^4},~~~
WeylSq=-\fr{6}{t^4}(-1-3p_3+4p_3^3),\n
&RiemSq=-\fr{3}{4t^4}(-24p_3+32p_3^3-17),
\ear
and
the {\it shear-free case}
when $a^2=b=3$, \ref{shears} shows there is vanishing shear scalar
and \ref{expsc} shows non-vanishing expansion,
this has curvature products
\be
R=\fr{6}{t^2},~~~
RicciSq=\fr{12}{t^4},~~~
WeylSq=0,~~~
RiemSq=\fr{12}{t^4}.
\ee
\section{Kasner spacetime in five dimensions.}
The line element is
\be
ds^2=-dt^2+t^{2p_1}dx^2+t^{2p_2}dy^2+t^{2p_3}dz^2+t^{2p_4}dw^2,
\ee
There is the traditional type solution,
with
\be
p_1+p_2+p_3+p_4=1,~~~
p_1^2+p_2^2+p_3^2+p_4^2=1,
\label{td5}
\ee
the Ricci, Pauli and Bach tensors all vanish.
The invariants are
\ber
&RiemSq=WeylSq=-\fr{8}{t^4}\times\\
&(-2p_3p_4+p_3^2p_4+p_3^3p_4+p_4^2p_3+p_4^3p_3+2p_3^3+2p_4^3-2p_3^2-2p_4^2+p_3^2p4^2).
\nonumber
\ear
For the constraints
\be
p_3=p_4=0,~~~
p_1+p_2=1
\label{gbd5}
\ee
the Gauss-Bonnet tensor vanishes
and the invariants are
\be
R=\fr{2p_2(p_2-1)}{t^2},~~~
RiemSq=3R^2,~~~
RicciSq=R^2,~~~
WeylSq=\fr{11}{6}R^2,
\ee
this is the same pattern as some scalar five dimensional solutions \cite{mdr}.
No other five dimensional Kasner solution has been found.
\newpage
\section{The Kasner one brane.}
The line element is
\be
ds^2=-\exp(\fr{\chi}{\sqrt{\alpha}})dt^2+d\chi^2,
\label{kb1}
\ee
this line element obeys the vacuum Einstein equations $G_{ab}=0$,
and has curvature invariants
\be
R=-\fr{1}{2\alpha},~~~
RiemSq=R^2,~~~
RicciSq=\fr{1}{2}R^2.
\ee
\section{The Kasner two brane.}
The line element is taken to be
\be
ds^2=\exp(\fr{\chi}{\sqrt{\alpha}})\left(-dt^2+t^{2p}dx^2\right)+d\chi^2.
\ee
After subtracting off the cosmological constant using
\be
\bar{R}_{ab}=R_{ab}+\frac{(d-1)}{4\alpha}g_{ab},~~~
\bar{G}_{ab}=G_{ab}+\frac{(d-1)(2-d)}{8\alpha}g_{ab},
\label{brg}
\ee
where $d$ is the dimension of the spacetime,  the curvature is given by
\be
\bar{R}_{tt}=V,~~~
\bar{R}_{xx}=-t^{2p}V,~~~
\bar{G}_{\chi\chi}=V\exp\left(-\frac{\chi}{\sqrt{\alpha}}\right),~~~
V\equiv\frac{p(p-1)}{t^2}.
\label{curvk2b}
\ee
The stress on the brane surface can be taken to be that of a $d=2$ scalar field in which the
derivatives of the field are negligible and the potential given by $V$ in \ref{curvk2b};
alternatively the stress can be taken to be that of a $d=3$ fluid with vector field
$U_a=(0,0,1),~\pr=0,~\mu=G_{\chi\chi}$,
this is less satisfactory as the fluid vector $U$ is not timelike nor in the brane.

No solutions for the `mixed up` brane line element
\be
ds^2=\exp(\fr{\chi}{\sqrt{\alpha}})\left(-dt^2+t^{2p_1}dx^2\right)+t^{2p_2}d\chi^2.
\ee
have been found for $p_2\ne0$.
\section{The Kasner three brane.}
The line element is taken to be
\be
ds^2=\exp(\fr{\chi}{\sqrt{\alpha}})\left(-dt^2+t^{2p_1}dx^2+t^{2p_2}dy^2\right)+d\chi^2.
\ee
After subtracting off the cosmological constant using \ref{brg}
the modified Ricci and Einstein tensors are
\ber
\bar{R}_{ab}&=&\frac{1}{t^2}{\rm diag}
\left[\sum p-\sum p^2,p_1(\sum p-1)t^{2p_1},p_2(\sum p-1)t^{2p_2},0\right],\\
\bar{G}_{ab}&=&\frac{1}{t^2}{\rm diag}
\left[p_1p_2,p_2(1-p_2)t^{2p_1},p_1(1-p_1)t^{2p_2},
(\sum p -\sum p^2-p_1p_2)\exp(-\frac{\chi}{\sqrt{\alpha}})\right],
\nonumber
\label{gbar3}
\ear
For $p_1=1,~p_2=0$ the line element has just the cosmological constant,
for $G_{\chi\chi}=0$ solving the quadratic gives
\be
p_2=\frac{1}{2}\left(1-p_1\pm\sqrt{(1-p_1)(1+3p_1)}\right),
\label{qu3}
\ee
if $p_1=p_2$ then \ref{qu3} gives both equal to $2/3$ and the modified stress \ref{gbar3}
is that of a perfect fluid \ref{pf} with
\be
U_a=\left[1,0,0\right],~~~~~~~
\mu=2\pr=\frac{4}{9t^2},
\label{3sol}
\ee
if $p_1\ne p_2$ then the spacetime cannot be that of a perfect fluid as a timelike
fluid vector $U$ cannot be chosen.

For $p_1=3,~p_2=0$ the vacuum-Bach equations are obeyed and the invariants are
\ber
&R=\fr{3}{\alpha t^2}\left(-t^2+8\alpha\exp(-\fr{\chi}{\sqrt{\alpha}}\right),~~~
WeylSq=192t^{-4}\exp(-\fr{2\chi}{\sqrt{\alpha}}),\n
&RiemSq=\fr{1}{6}R^2+\fr{5}{2}WeylSq,~~~
RicciSq=\fr{1}{4}R^2+\fr{3}{4}WeylSq.
\ear
\section{The Kasner brane fluid.}
The line element is taken to be
\be
ds_5^2=\exp\left({\fr{\chi}{\sqrt{\alpha}}}\right)ds_{d={\rm 4kasner}}^2+d\chi^2,
\label{kasnerd5}
\ee
with $ds^2_{d={\rm 4kasner}}$ given by \ref{kasnerle}.
The metric is not Ricci-flat for any values of the $p$'s
as the $\exp(\chi/\sqrt{\alpha})$ term always occurs.
After subtracting off the cosmological constant using \ref{brg}
and using the notation \ref{ab} the modified Ricci and Einstein tensors are
\ber
\bar{R}_{ab}&=&\frac{1}{t^2}\left[b-a^2,t^{2p_i}p_i(b-1),0\right],\\
\bar{G}_{ab}&=&\frac{1}{t^2}
\left[c,b-p_i-a^2+p_i^2,(b-a^2-c)\exp\left(-\frac{\chi}{\sqrt{\alpha}}\right)\right],
\nonumber
\label{4rg}
\ear
where $i$ is not summed;  as for the three brane \ref{3sol} take the $p$'s all equal,
the value which gives $\bar{G}_{\chi\chi}=0$ is $p=1/2$ and the result is a perfect fluid
with
\be
U_a=[1,0,0,0],~~~
\mu=3\pr=\frac{3}{4t^2},
\label{4sol}
\ee
which has a radiation equation of state.
Transferring the $d=4$ solution to spherical coordinates
\be
ds^2=-dt^2+t~d\Sigma_3^2,~~~
\Sigma_3^2=dr^2+r^2(d\theta^2+\sin(\theta)^2d\phi^2),
\label{sp4sol}
\ee
grtensor calculates the Ricciscalar,  Weyl, Bach and Pauli tensors all to vanish
and the non-vanishing curvature is most easily expressed
in terms of the Ricci scalar $\Phi_{00}$
\ber
&&\Phi_{00}=4\Phi_{11}=4\Phi_{22}=\frac{1}{2t^2},\\
&&RiemSq=6\Phi_{00},~~~
RicciSq=3\Phi_{00},\n
&&R_1=\frac{3}{4}\Phi_{00}^2,~~~
R_2=\frac{3}{8}\Phi_{00}^3,~~~
R_3=\frac{21}{64}\Phi_{00}^4.
\nonumber
\label{sol4invars}
\ear
\section{The quadratic order Kasner four brane.}
For the line element \ref{kasnerd5} using the traditional constraints \ref{ab} with $a,b=1$,
the curvature products are
\ber
&R=-\fr{5}{\alpha},~~~
RicciSq=\fr{1}{5}R^2,\\
&WeylSq=\fr{16p_3^2(1-p_3)}{t^4}\exp(2\fr{\chi}{\sqrt{\alpha}}),~~~
RiemSq=WeylSq+\fr{1}{10}R^2.\nonumber
\ear
So far no solutions involving the Bach tensor have been found.
The field equations
\be
G_{ab}-3P_{ab}/5=0,
\label{ped5}
\ee
and
\be
G_{ab}+\alpha GB_{ab}+\alpha\delta^{\chi\chi}_{ab}WeylSq=0,
\label{gbe}
\ee
are obeyed.
For the expansion-free constraints $b=0~\&~2a^2=3$
and the shear-free constraints $a^2=b=3$ the Bach tensor is non-vanishing,
but no field equations have been found to be obeyed.
\section{The double Kasner solution.}\label{mixed}
Instead of choosing a five space of constant curvature it is possible to choose the five
space to be of Kasner form.  An example of this is the double Kasner metric
\be
ds_5^2=\chi^{2q_1}t^{2p_1}dx^2+\chi^{2q_2}t^{2p_2}dy^2+\chi^{2q_3}t^{2p_3}dz^2
-\chi^{2q_4}t^{2p_4}dt^2+\chi^{2q_5}t^{2p_5}d\chi^2,
\label{doublekasner}
\ee
which has Ricci tensor
\ber
R_i^{~i}&=&\chi^{-2q_4}t^{-2p_4-2}p_i\left(P_5-2p_4-1\right)
          -\chi^{-2q_5-2}t^{-2p_5}q_i\left(Q_5-2q_5-1\right),\n
R_t^{~t}&=&\chi^{-2q_4}t^{-2p_4-2}\left(P_5^2-p_4^2+(p_4-P_5)(1+p_4)\right)
          -\chi^{-2q_5-2}t^{-2p_5}q_4\left(Q_5-2q_5-1\right),\n
R_\chi^{~\chi}&=&\chi^{-2q_5-2}t^{-2p_5}\left(Q_4(1+q_5)-Q_4^2\right)
                 -\chi^{-2q_4}t^{-2p_4-2}\left(-P_5+2p_4+1\right),\n
R_\chi^{~t}&=&\chi^{-2q_4-1}t^{-2p_4-1}\left(PQ-q_4P_3-p_5Q_3\right),
\label{dkricci}
\ear
where
\be
PQ\equiv\sum_{i=1}^3q_ip^i,~~~
P_j\equiv\sum_{i=1}^jp_i,~~~
P_j^2\equiv\sum_{i=1}^jp_i^2,
\ee
and similarly for $q$.
There is the non-interacting vacuum solution
\be
q_1=q_2=p_2=p_3=\frac{2}{3},~~~
p_1=q_3=-\frac{1}{3},~~~
p_4=p_5=q_4=q_5=0.
\label{noninter}
\ee
There is no immediate interacting vacuum solution to these constraints,  as for example
\be
p_1=p_2=\frac{2}{3},~
p_3=-\frac{1}{3},~~
q_1=q_2=q_3=\frac{1}{2},~
q_4=-\frac{1}{2},~~
p_4=p_5=q_5=0,
\label{dkcon}
\ee
solves the non-interacting equations but gives the wrong sign for the interacting $PQ$ equation.
The choice
\be
p_1=p_2=q_1=q_2=\frac{2}{3},~
p_3=q_3-\frac{1}{3},~
p_4=-q_4=-a,~
p_5=-q_5=1-a,~
\ee
where $a$ is a constant, is a vacuum solution with non-vanishing Riemann components;
however the Kretschmann invariant vanishes $RiemSq=K=0$, also the quadratic tensors vanish.
\section{Matter trapping}\label{mt}
In order to consider matter trapping \cite{RuS,SV} one takes test objects,
such as particles,  fields or fluids,  and investigates whether they coalesce toward the brane
or disperse.  A criteria to judge whether this is happening whether the energy conditions
\cite{HE}\S4.3 or equations similar to them are obeyed.   For example a minimal scalar field
obeys the null converges condition,  but this is not sufficient for minimal scalar fields always
to coalesce toward the brane because the energy condition for the brane spacetime depends on its
Ricci tensor rather the test minimal scalar's hatted Ricci tensor.
Consider a massless variable separable Klein-Gordon test scalar field in the spacetime
with line element given by \ref{kasnerd5} and \ref{kasnerle},  a solution is
\be
\psi=At^{(1-b)}{\rm erf}\left(\alpha^{-\frac{1}{4}}\chi\right),
\label{psikg}
\ee
where $A$ is a constant and $b$ given by \ref{ab}.
For the null vector
\be
N^a=\exp\left(-\frac{\chi}{\sqrt{\alpha}}\right)\delta^a_t\pm\delta^a_\chi
\label{nuv}
\ee
the null convergence condition is
\ber
N^aN^b\hat{R}_{ab}&=&2\left\{\exp\left(-\frac{\chi}{\sqrt{\alpha}}\right)\psi_t\pm\psi_\chi\right\}^2\\
&=&\frac{8A^2t^{-2b}}{\pi\sqrt{\alpha}}\left\{\pm t+(1-b)\chi+\Order\left(\chi^2\right)\right\}^2,
\nonumber
\label{nnr}
\ear
as this is positive this particular Klein-Gordon matter converges toward the brane.
It is not immediate what sort of energy conditions and vector fields are best in the non-null case,
in particular should the vector field $V$ be timelike or in some sense point toward the brane,
and the usual energy conditions hold for an arbitrary timelike vector field,  should this arbitrariness
still be present or could a specific vector field be used.
As a second test object consider a perfect fluid \ref{pf} with timelike vector field,
acceleration and expansion
\be
V_a=\exp\left(\frac{\chi}{2\sqrt{\chi}}\right)\delta^t_a,~~~
\dot{V}^a=\frac{1}{2\sqrt{\alpha}}\delta^a_\chi,~~~
\st{V}{\theta}=-\frac{b}{t}\exp\left(-\frac{\chi}{2\sqrt{\alpha}}\right),
\label{vecpf}
\ee
where $b$ is given by \ref{ab}.  The conservation equations are
\be
T^{~b}_{t.;b}=-\mu_t-\frac{b}{t}\left(\mu+\pr\right),~~~
T^{~b}_{\chi.;b}=\pr_\chi+\frac{1}{2\sqrt{\alpha}}\left(\mu+\pr\right),
\label{testce}
\ee
assuming the equation of state \ref{geqs} and separation of variables gives the solution
\be
\mu=At^{-b\gamma}\exp\left(-\frac{\gamma\chi}{2(\gamma-1)\sqrt{\alpha}}\right),
\label{psol}
\ee
where $A$ is a constant:  this solution decays for large $\chi$.
\section{Conclusion.}\label{conc}
For higher order tensors.
The $d=2$ Kasner spacetime \ref{pd2} has a vacuum-Pauli solution when $p=-3$.
No solutions to $d=3$ Kasner spacetime have been found.
The $d=4$ Kasner spacetime is a vacuum-Einstein solution when the $p$'s obey the
traditional constraints,  however there are additional vacuum-Bach solutions
when the $p$'s obey expansion-free or shear-free constraints \ref{bsq}.
The $d=5$ Kasner spacetime has a solution similar to the traditional $d=4$ solution \ref{td5},
and also a simple solution which has vanishing Gauss-Bonnet tensor \ref{gbd5}.
The Kasner one brane \ref{kb1} is a solution to the vacuum-Einstein equations.
The Kasner two brane is a solution to the Pauli-Einstein equations when $p=1$.
The Kasner three brane is a solution when $p_1=p,~p_2=0$,
for $p=+1$ it is a solution to both the vacuum-Pauli and vacuum-Bach equations,
for $p=-3$ it is a solution to the vacuum-Bach equations.
The Kasner four brane has a solutions to the
Pauli-Einstein \ref{ped5} and also to the Gauss-Bonnet-Einstein equations \ref{gbe}.

For double Kasner spacetime \ref{doublekasner}
a non-interacting vacuum solution was found \ref{noninter},
but no interacting solution was found.

For brane fluids,  with a perfect fluid in the brane and a vacuum in the bulk
three brane \ref{3sol} and four brane \ref{4sol} solutions were found.
The four brane solution has a four metric the same as $k=0$ radiation
filled Robertson-Walker spacetime.

For brane fluids, a test scalar field obeys a null convergence condition \ref{nnr},
and in this sense the field converges onto the brane; a test perfect fluid \ref{psol}
decays away from the brane and in this sense is trapped on the brane.
\newpage

\end{document}